\documentclass[12pt]{article}
\usepackage{amsmath,epsf,amssymb,latexsym,cite}
\usepackage[ps,dvips,matrix,arrow,frame,import,curve,color]{xy}

\makeatletter
\renewcommand\section{\@startsection {section}{1}{\z@}%
                                   {-3.5ex \@plus -1ex \@minus -.2ex}%
                                   {2.3ex \@plus.2ex}%
                                   {\normalfont\bfseries}}
\makeatother

\newtheorem{theorem}{Theorem}

\newif\iffigs\figstrue

\DeclareFontFamily{U}{rsf}{}
\DeclareFontShape{U}{rsf}{m}{n}{
  <5> <6> rsfs5 <7> <8> <9> rsfs7 <10-> rsfs10}{}
\DeclareMathAlphabet\Scr{U}{rsf}{m}{n}

\def\pplogo{\vbox{\kern-\headheight\kern -29pt
\halign{##&##\hfil\cr&{
\ppnumber}\cr\rule{0pt}{2.5ex}&\ppdate\cr}
}}
\makeatletter
\def\ps@firstpage{\ps@empty \def\@oddhead{\hss\pplogo}%
  \let\@evenhead\@oddhead 
}
\def\maketitle{\par
 \begingroup
 \def\thefootnote{\fnsymbol{footnote}}
 \def\@makefnmark{\hbox{$^{\@thefnmark}$\hss}}
 \if@twocolumn
 \twocolumn[\@maketitle]
 \else \newpage
 \global\@topnum\z@ \@maketitle \fi\thispagestyle{firstpage}\@thanks
 \endgroup
 \setcounter{footnote}{0}
 \let\maketitle\relax
 \let\@maketitle\relax
 \gdef\@thanks{}\gdef\@author{}\gdef\@title{}\let\thanks\relax}
\makeatother

\def\O{\Scr{O}}
\def\C{{\mathbb C}}
\def\P{{\mathbb P}}

\def\R{{\mathbb R}}
\def\Z{{\mathbb Z}}

\def\Im{\operatorname{Im}}
\def\Hom{\operatorname{Hom}}

\def\Ext{\operatorname{Ext}}
\def\End{\operatorname{End}}

\def\coker{\operatorname{coker}}

\def\Gl{\operatorname{GL}}

\def\SU{\operatorname{SU}}

\def\rank{\operatorname{rank}}

\def\CY{Calabi--Yau}

\def\cE{{\Scr E}}

\def\DC{\mathbf{D}}

\def\dP#1{\mathrm{dP}_{#1}}
\def\cmod#1{\hbox{$#1$--\bf mod}}
\def\obj#1{\mathsf{\lowercase{#1}}}
\def\poso#1{#1\save="x"!LD+<0pt,-0.5mm>;
  "x"!RD+<0pt,-0.5mm>**\dir{.}\restore}

\begin{document}
\setcounter{page}0
\def\ppnumber{\vbox{\baselineskip14pt
\hbox{DUKE-CGTP-04-08}
\hbox{hep-th/0407123}}}
\def\ppdate{July 2004} \date{}

\title{\LARGE D-Branes, $\Pi$-Stability and $\theta$-Stability\\[10mm]}
\author{
Paul S.~Aspinwall\\[2mm]
\normalsize Center for Geometry and Theoretical Physics \\
\normalsize Box 90318 \\
\normalsize Duke University \\
\normalsize Durham, NC 27708-0318
}

{\hfuzz=10cm\maketitle}

\def\Large{\large}
\def\LARGE{\large\bf}

\vskip 1cm

\begin{abstract}
We investigate some aspects of $\Pi$-stability of D-branes on \CY\
threefolds in cases where there is a point in moduli space where the
grades nearly or completely align. We prove that an example of
complete alignment is the case of a collapsed del Pezzo surface. It is
shown that there is an open neighbourhood of such a point for which
$\Pi$-stability reduces to $\theta$-stability of quiver
representations. This should be contrasted to the case of the large
radius limit where $\mu$-stability of sheaves cannot be extended out
over an open neighbourhood.
\end{abstract}

\vfil\break


\section{Introduction}    \label{s:intro}

$\Pi$-stability of B-type BPS D-branes on a \CY\ threefolds $X$
captures many interesting aspects of ``stringy geometry''. At large
radius limit such D-branes correspond to Hermitian-Yang--Mills
connections on bundles over holomorphically embedded subspaces in
$X$. The instant one moves away from this limit, nonperturbative
string tension (i.e., $\alpha'$) corrections change this picture quite
drastically. One must pass to the bounded derived category $\DC(X)$ of
coherent sheaves to understand all physical D-branes.

Objects in the derived category are, roughly speaking, equivalence
classes of complexes of coherent sheaves. A complex which is nonzero
in only one position is called a ``stalk complex''. If an object in
$\DC(X)$ can be represented by a stalk complex then we may regard this
object as simpler in some sense. Indeed, the classical D-branes at
large radius limit correspond to stalk complexes whose nonzero entry
is a coherent sheaf supported on a subspace of $X$.

It was shown in \cite{Doug:DC,Doug:S01,AD:Dstab} that for arbitrarily large but
finite $X$, one may always find a $\Pi$-stable object in $\DC(X)$ that
is not isomorphic to a stalk complex. Therefore, any method of
analyzing B-type D-branes near the large radius limit in terms of bundles
and connections (such as in \cite{MMMS:BPS,MT:mu-stab,EL:mupi}) is
always fated to miss some of the physical objects.

The category of B-type D-branes does not ``know'' it is associated
specifically with complexes of coherent sheaves. It just happens that
coherent sheaves become important at large radius. There may be other
``abelian categories'' which, when ``derived'', correctly yield the
same D-brane category in passing to complexes.

There are other points in the moduli space where another abelian
category becomes useful. Of interest to us in this lecture is the case
where the ``gradings'' of the stable D-branes within this abelian
category lie in a narrow range as spelt out more precisely below. In
this case we will have an open neighbourhood of the moduli space where
all stable objects can be given by stalk complexes. Our ideas are not
particularly new. Much of what we discuss has been observed in
\cite{DFR:stab,Doug:DC}, for example, and is close to some ideas of
Bridgeland \cite{Brg:stab}. Our goal here is to present a systematic
approach to the problem which lets us arrive at some theorems which
clarify the generality of the situation.

We shall prove that this ``alignment'' of gradings happens if a del
Pezzo surface within $X$ is shrunk. That is, when the del Pezzo
surface is completely contracted to a point, the grades will align
completely. If $S$ is shrunk down to something small (relative to the
intrinsic $\alpha'$ scale) the grades will align sufficiently, so that
all $\Pi$-stable objects are stalk complexes. In this open region of
moduli space one may then use a simpler notion of stability, namely
$\theta$-stability, to accurately determine the stable spectrum.

Such a simplification appears to be very special. As mentioned above,
$\mu$-stability (or probably any variant thereof) for vector bundles
is not valid in an open set in moduli space. Similarly, the gradings
of stable objects in Gepner models do not appear to align in any way,
and so we should not expect any such simplification here either.

The derived category in question will be based on quiver
representations rather than coherent sheaves. Quivers have played an
important r\^ole in the theory of D-branes in the context of quiver
gauge theories and del Pezzo surfaces as explored in
\cite{DM:qiv,MP:AdS,BGLP:dPo,BP:toric,FHHI:quiv,HI:quiv,CFIKV:,
Wijn:dP,HW:dib,Herz:exc,Herz:ouch,AM:delP} for example. We will not
pursue these quiver gauge theories in this paper except to say that we
prove that these theories become appropriate precisely when the del
Pezzo surface shrinks down to a point as had been suspected.

In section \ref{s:quiv} we review the necessary concepts from quivers
and their representations. Then, in section \ref{sec:pts}, we focus on
D-branes that correspond to points in $S$. By looking at the
moduli space of such objects we retrieve $S$ itself. In section
\ref{s:theta} we analyze the case where the gradings are partially
aligned and we prove that $\Pi$-stability reduces to
$\theta$-stability in this case. Finally in section \ref{s:align} we
consider a complete alignment of the gradings and show that the del
Pezzo surface collapses in this case.


\section{Quivers and del Pezzo Surfaces}    \label{s:quiv}

This section is a quick review of some well-known material. See
\cite{Ben:quiv,AM:delP}, for example, for more details. Let $Q$ be a
quiver. The path algebra $A$ (over $\C$) of $Q$ is constructed as
follows. To each node of $Q$ we associate a element $e_j\in A$ which
is idempotent, $e_j^2=e_j$. These elements are viewed as ``paths of
zero length'' beginning and ending at node $j$. The rest of $A$ is
then generated by paths of nonzero length in $Q$. If the end of path
$p$ is the same node as the beginning of path $q$, then the product
$qp$ is simply the composition of the paths. Otherwise the product is
zero. A quiver may contain {\em relations\/} imposed between the
paths. 

A representation of $Q$ is defined by associating a vector space $V_j$
over $\C$ with each node. In addition, a matrix yielding a linear map
is then associated to each arrow. By using matrix multiplication we
thus associate a matrix to any path in $Q$. If $Q$ has relations, the
matrices must obey these relations.

It is not hard to show (see \cite{Ben:quiv}, for example) that quiver
representations are equivalent to representations of $A$, or left
$A$-modules. If $V$ is a vector space and a left $A$-module, one sets
$V_j=e_j V$.  The {\em dimension vector\/} of a quiver representation
is simply the list of numbers $(\dim V_0,\dim V_1,\ldots)$.

A morphism between two quiver representations can then be defined to
correspond to an $A$-module homomorphism. This means that, given two
representations $W$ and $V$ with underlying vector spaces $W_j$ and
$V_j$ at each node, we must specify a collection of linear maps
\begin{equation}
  f_j:W_j\to V_j,
\end{equation}
such that for any arrow $a$,
\begin{equation}
  f_{h(a)}W_a = V_af_{t(a)},
\end{equation}
where $V_a$ is the matrix $V$ associates to $a$, and $h(a)$ and $t(a)$
are the head and tail of $a$ respectively. From these morphisms we
define the category of quiver representations which is equivalent to
the category of left $A$-modules.

There are two particularly useful sets of quiver representations,
$L_j$ and $P_j$, each of which is labeled by a node $j$ in the
quiver. $L_j$ is defined simply as the one-dimensional representation
with dimension vector $(0,\ldots,0,1,0,\ldots)$ with the one in the
$j$th position.

$P_j$ is defined as $Ae_j$, i.e., the space generated by paths
starting at node $j$. It is clear that $\sum_j e_j$ is the identity
element in $A$, and thus
\begin{equation}
\begin{split}
A &= A\sum_j e_j\\
  &= \bigoplus_j P_j.
\end{split}
\end{equation}
By a standard theorem of algebra, each $P_j$ is therefore a projective
$A$-module since it is a direct summand of $A$ (which is obviously a
free $A$-module).

We may now construct the derived category $\DC(\cmod A)$, which we
will also denote $\DC(Q)$, from bounded complexes of quiver
representations in the usual way (see \cite{me:TASI-D} for a review of
this). A ``stalk complex'' is a complex which is nonzero in precisely
one position. As always $\obj A[n]$ denotes the complex $\obj A$
shifted $n$ places to the left. When used as an object of $\DC(\cmod A)$,
$V$ will represent a stalk complex with quiver representation $V$ in position
zero. Similarly, $V[n]$ represents a stalk complex with $V$ in
position $-n$.

Suppose $S$ is a del Pezzo surface and $\{\cE_0,\ldots,\cE_{n-1}\}$
forms a strong exceptional collection of sheaves on $S$ as in
\cite{Rud:Ebook}. That is,
\begin{equation}
\begin{split}
\Ext_S^q(\cE_j,\cE_j) &= \begin{cases}
  \C&\text{if $q=0$},\\
  0&\text{otherwise}, \end{cases}\\
\Ext_S^q(\cE_j,\cE_k) &= 0 \quad\text{for any $q$ and $j>k$}.
\end{split}
\end{equation}
Furthermore, assume that the number of sheaves in this collection,
$n$, is equal to the rank of $K_0(S)$. Bondal \cite{Bon:dPq} then
proved that
\begin{theorem} \label{th:equiv}
  The bounded derived category of coherent sheaves on $S$, $\DC(S)$, is
  equivalent to $\DC(\cmod A)$, the bounded derived category of left
  $A$-modules, where
\begin{equation}
  A = \End(\cE_0\oplus\ldots\oplus\cE_{n-1})^{\mathrm{op}}.
\end{equation}
\end{theorem}
Given such an exceptional collection it is easy to construct the
quiver $Q$ for which $A$ is the path algebra. There will always be
relations. For example, consider $S=\dP1$ given by $\P^2$ with the
single point $[z_0,z_1,z_2]=[0,0,1]$ blown up. Let $C_1$ be the
exceptional curve and let $H$ be a hyperplane not intersecting $C_1$.
Using the strongly exceptional collection
$\{\O,\O(C_1),\O(H),\O(2H)\}$, the corresponding quiver is given by
\begin{equation}
\begin{xy} <1.0mm,0mm>:
  (0,0)*{\circ}="a",(20,0)*{\circ}="b",(40,0)*{\circ}="c",(60,0)*{\circ}="d",
  (0,-4)*{v_0},(20,-4)*{v_1},(40,-4)*{v_2},(60,-4)*{v_3}
  \ar@{<-}|{a} "a";"b"
  \ar@{<-}@/^3mm/|{b_0} "b";"c"
  \ar@{<-}|{b_1} "b";"c"
  \ar@{<-}@/_8mm/|{c} "a";"c"
  \ar@{<-}@/^3mm/|{d_0} "c";"d"
  \ar@{<-}|{d_1} "c";"d"
  \ar@{<-}@/_3mm/|{d_2} "c";"d"
\end{xy}  \label{eq:dP1}
\end{equation}
subject to the relations $b_0d_1-b_1d_0=0$, $ab_0d_2-cd_0=0$, and
$ab_1d_2-cd_1=0$.

Given that $A=\End(A)^{\mathrm{op}}=\End(\oplus_j P_j)^{\mathrm{op}}$,
there is a natural identification of the exceptional sheaves $\cE_j$
with the projective representations $P_j$ which yields the equivalence
of theorem \ref{th:equiv}. Note that the definition of the exceptional
collection implies that there are no paths in $Q$ from node $i$ to
node $j$ if $i<j$. This means the quiver is ``directed'' and contains
no oriented loops. This, in turn, implies that $A$, and thus all the
$P_j$'s, are finite-dimensional.

Another consequence of the fact that the quiver is directed is that we
may decompose any quiver representation $E$ into a sum of $L_j$'s via
a sequence of short exact sequences. We can represent this by the
following triangles in $\DC(S)$:
\begin{equation}
\xymatrix@!C=3.5mm{
**[l]L_{0}^{\oplus N_{0}}=E_{0}\ar[rr]&&E_{1}\ar[dl]
\ar[rr]&&\cdots\ar[dl]\ar[rr]&&E_{n-2}\ar[dl]
\ar[rr]&&**[r]E_{n-1}=E\ar[dl]\\
&L_{1}^{\oplus N_{1}}\ar[ul]|{[1]}&&L_{2}^{\oplus N_{2}}\ar[ul]|{[1]}
&\ldots&L_{n-2}^{\oplus N_{n-2}}\ar[ul]|{[1]}&&L_{n-1}^{\oplus N_{n-1}}
\ar[ul]|{[1]}
} \label{eq:bchain}
\end{equation}
In this way, the $L_j$'s are the ``fundamental'' representations into
which any representation may be decomposed in the sense of
(\ref{eq:bchain}).

Viewing the $L_j$'s as D-branes, any representation may be interpreted
as a quiver gauge theory in which the field content is given by open
strings, i.e., $\Ext^q(L_j,L_k)$. The details of this quiver gauge
theory are not important for us here and we refer to \cite{AM:delP}
and references therein for more information for the interested
reader. An important condition for this quiver gauge theory to be
physically meaningful is that \cite{AM:delP,Herz:ouch} 
\begin{equation}
  \Ext^q(L_j,L_k) = 0, \qquad\hbox{for $q\neq1$ or 2, and any $j,k$.}
\end{equation}
This condition amounts to there being no relations between relations
within the quiver. We will assume $Q$ satisfies this condition from
now on.

Suppose we have an embedding $i:S\hookrightarrow X$ of our del Pezzo
surface $S$ into a \CY\ threefold $X$. The induced map $i_*:\DC(S)\to
\DC(X)$ maps $\DC(S)$ into a subcategory of $\DC(X)$ but this is not
{\em full\/} subcategory. That is, for quiver representations $A$ and
$B$ it need not be true that $\Ext_S^q(A,B)$ is equal to
$\Ext^q_X(i_*A,i_*B)$. We may remedy this by adding more arrows to the
quiver $Q$ as follows. First we note that a spectral sequence
associated to the embedding of $S$ into $X$ leads to the following
relation:
\begin{equation}
  \Ext^q_X(i_*A,i_*B) = \Ext^q_S(A,B) \oplus \Ext^{3-q}_S(B,A).
\end{equation}
In particular,
\begin{equation}
  \Ext^1_X(i_*L_j,i_*L_k) = \Ext^1(L_j,L_k) \oplus \Ext^2(L_k,L_j).
\end{equation}
Since the number of arrows from node $j$ to node $k$ is given by the
dimension of $\Ext^1(L_j,L_k)$ (see, \cite{me:TASI-D}, for example for
an explanation of this), we may create a ``completed'' quiver $\bar Q$
which yields the correct values of $\Ext^1_X(i_*L_j,i_*L_k)$ by adding
$\dim\Ext^2(L_k,L_j)$ arrows from node $j$ to node $k$. One can show
that $\dim\Ext^2(L_k,L_j)$ is precisely the number of independent
relations between paths from node $k$ to node $j$. Our example of
$\dP1$ above therefore becomes:
\begin{equation}
\begin{xy} <1.0mm,0mm>:
  (0,0)*{\circ}="a",(20,0)*{\circ}="b",(40,0)*{\circ}="c",(60,0)*{\circ}="d",
  (0,-4)*{v_0},(20,-4)*{v_1},(40,-4)*{v_2},(60,-4)*{v_3}
  \ar@{-}|*\dir{<}"a";"b"
  \ar@{-}@/_1mm/|*\dir{<}"b";"c"
  \ar@{-}|*\dir{<}"b";"c"
  \ar@{-}@/_6.5mm/|(0.3)*\dir{<}"a";"c"
  \ar@{-}@/_1mm/|*\dir{<}"c";"d"
  \ar@{-}|*\dir{<}"c";"d"
  \ar@{-}@/^1mm/|*\dir{<}"c";"d"
  \ar@{.}@/^6mm/|*\dir{>}"a";"d"
  \ar@{.}@/^5mm/|*\dir{>}"a";"d"
  \ar@{.}@/^4mm/|(0.3)*\dir{>}"b";"d"
\end{xy}  \label{eq:dP1X}
\end{equation}
We will always use dotted arrows to represent the new arrows added in.

Once we have the correct value for the $\Ext^1$'s between the $L_j$'s,
it follows from Serre duality $\Ext^q_X(A,B)=\Ext^{3-q}_X(B,A)$, that
the $\Ext^2$'s are also correct. Since all other $\Ext$'s vanish, this
quiver also yields the correct value for all $\Ext$'s  between the
$L_j$'s. Furthermore, since we can decompose any quiver representation
(and thus any object in $\DC(S)$) into $L_j$'s, it follows that the
quiver $\bar Q$ yields the correct values for all the $\Ext$'s between
all objects. In other words, if $\bar A$ is the path algebra of the
completed quiver $\bar Q$, $\DC(\cmod{\bar A})$ maps to a {\em full\/} 
subcategory of $\DC(X)$.

The category $\DC(\cmod{\bar A})$, which we will also denote $\DC(\bar
Q)$, represents local information of $X$ near $S$. As we will see, it
appears to represent the derived category of coherent sheaves with
compact support on the total space of the normal bundle of $S$ in
$X$. We will not attempt to prove this rigorously here though.


\section{Points and Stability} \label{sec:pts}

In this section we focus attention on the objects in $\DC(S)$ which
correspond to points, i.e., skyscraper sheaves $\O_p$, $p\in S$. In
\cite{BO:DCeq} Bondal and Orlov gave a purely categorical method of
determining which elements of $\DC(S)$ correspond to such
skyscrapers. They showed
\begin{theorem} \label{th:BO}
  Assuming $S$ is a smooth del Pezzo surface, an object $\obj A$ in
$\DC(S)$ corresponds to a stalk complex given by a skyscraper sheaf if
and only if the following three conditions hold
\begin{enumerate}
\item $\Psi(\obj A) = \obj A[2]$,
\item $\Hom(\obj A,\obj A[n]) = 0$, for $n<0$,
\item $\Hom(\obj A,\obj A)=\C$.
\end{enumerate} 
\end{theorem}
Here $\Psi$ is the ``Serre-functor'' \cite{BK:Serre} which is unique
up to isomorphism and, for derived categories of quiver
representations, is given by the ``Nakayama functor'':\footnote{$A$ is
a left-right-$A$-module. Therefore, if $V$ is a left $A$-module,
$\Hom(V,A)$ is a right $A$-module. Taking the dual of this restores to
us a left $A$-module.}
\begin{equation}
  \Psi(\obj A) = \mathbf{R}\Hom(\obj A,A)^*.
\end{equation}

We will assume that $\O_p$ is not only a stalk complex in $\DC(S)$,
but is also a stalk complex when viewed as an object in the derived
category of quiver representations. This assumption is discussed
further in section \ref{s:theta}. It was proven in \cite{AM:delP} that
the dimension vector of a quiver representation corresponding to
$\O_p$ is given by
\begin{equation}
(\rank{\cE_0},\rank{\cE_1},\ldots,\rank{\cE_{n-1}}).  \label{eq:skydim}
\end{equation}
This is a weaker constraint than theorem~\ref{th:BO}.

Let us analyze the example of $\dP1$ and the quiver given in
(\ref{eq:dP1}). The quiver representation corresponding to $\O_p$ will
have a dimension vector $(1,1,1,1)$. Note that
\begin{equation}
\begin{split}
  e_j\Psi(\obj A) &= \mathbf{R}\Hom(\obj A,Ae_j)^*\\
     &= \mathbf{R}\Hom(\obj A,P_j)^*.
\end{split}
\end{equation}
Suppose $V$ is a quiver representation with dimension $(1,1,1,1)$ with
{\em generic\/} maps associated to the arrows. This has a projective
resolution
\begin{equation}
\xymatrix@1{
0\ar[r]&P_0\ar[r]&P_2^{\oplus2}\ar[r]&P_3\ar[r]&V\ar[r]&0.
}  \label{eq:dP1p}
\end{equation}
In other words, in $\DC(Q)$, $V$ is isomorphic to the complex
\begin{equation}
\xymatrix@1{
P_0\ar[r]&P_2^{\oplus2}\ar[r]&\poso{P_3},
}
\end{equation}
where the underline represents position zero. So
\begin{equation}
\begin{split}
  e_0\Psi(V) &= \mathbf{R}\Hom(\xymatrix@1{
P_0\ar[r]&P_2^{\oplus2}\ar[r]&\poso{P_3}
},P_0)^*\\
  &= \left(\xymatrix@1{
\C\ar[r]&0\ar[r]&\poso{0}}\right)\\
  &= \C[2].
\end{split}
\end{equation}
Similarly $e_j\Psi(V)= \C[2]$ for $j=1,2,3$. Thus the correspondence
of $V$ to $\O_p$ is consistent with the first condition in
theorem~\ref{th:BO}. It is not hard to show that the other two
conditions are also satisfied. However, suppose we impose $b_0=b_1=0$
on the maps in (\ref{eq:dP1}). Now $P_3\to V$ is no longer a
surjective map and so $V$ becomes isomorphic to the complex
\begin{equation}
\xymatrix@1{
P_0\ar[r]&P_1\oplus P_2^{\oplus2}\ar[r]^{(0,-)}&\poso{P_1\oplus P_3},
} \label{eq:dP1pf}
\end{equation}
which, in turn, yields
\begin{equation}
  e_1\Psi(V) = \left(\xymatrix@1{
\C\ar[r]&\C\ar[r]^0&\poso{\C}}\right),
\end{equation}
which is {\em not\/} isomorphic to $\C[2]$. Thus $b_0=b_1=0$ cannot
correspond to a skyscraper sheaf $\O_p$ for any $p$. The general rule
is that there must be a nonzero path of nonzero length to each
node. We therefore also rule out $a=c=0$ and $d_0=d_1=d_2=0$.

In this case it is easy to see $\dP1$ as the moduli space of
isomorphism classes of all valid $V$'s. First note that two quiver
representations are isomorphic if and only the map between the two
representations is an element of the ``gauge group''
\begin{equation}
  \prod_j \Gl(V_j).
\end{equation}
This means we have an overall $(\C^*)^4$ gauge group acting on our
representations, although, as always, a diagonal $\C^*$ acts
trivially. Suppose $d_0$ and $d_1$ are assigned values not both
zero. The remaining $(\C^*)^3$ action together with the quiver
relations and above non-vanishing conditions are then sufficient to
fix $a,b_0,b_1$ and $c$. If $d_0=d_1=0$ then $a=0$, and $c$ and $d_2$
are fixed, but the ratio of $b_0$ and $b_1$ remains undetermined. This
explicitly realizes $[d_0,d_1,d_2]$ as the homogeneous coordinates of
the original $\P^2$ and $[b_0,b_1]$ as the homogeneous coordinates of
the exceptional $\P^1$ when $d_0=d_1=0$.


\section{$\theta$-stability}  \label{s:theta}

The key step in obtaining $S$ as the moduli space of quiver
representations in section \ref{sec:pts} is using the criteria of
theorem \ref{th:BO} to rule out some representations as valid
skyscraper shaves. There is an alternative way of doing this. This
concerns the idea of {\em stability\/} either in the mathematical
sense or the physical sense. 

If these quiver representations are to correspond to physical
0-branes, they must be $\Pi$-stable in the sense of
\cite{DFR:stab,Doug:DC,AD:Dstab,Brg:stab,me:TASI-D}.  In this picture
one puts a stability condition on $\DC(X)$ as follows. Each nonzero
object in $\DC(X)$ is either stable or unstable. The stable objects
$\obj A$ have ``grades'' $\xi(\obj A)\in\R$ such that $\xi(\obj
A[n])=\xi(\obj A)+n$ and any two stable objects satisfy the unitarity
constraint
\begin{equation}
  \xi(\obj A)>\xi(\obj B)\quad\Rightarrow\quad\Hom(\obj A,\obj B)=0.
      \label{eq:unt}
\end{equation}
Every unstable object $\obj C$ then has a ``decay chain'' given by a sequence
of distinguished triangles
\begin{equation}
\xymatrix@!C=3.5mm{
**[l]\obj A_{0}=\obj C_{0}\ar[rr]&&\obj C_{1}\ar[dl]
\ar[rr]&&\cdots\ar[dl]\ar[rr]&&\obj C_{n-2}\ar[dl]
\ar[rr]&&**[r]\obj C_{n-1}=\obj C,\ar[dl]\\
&\obj A_{1}\ar[ul]|{[1]}&&\obj A_{2}\ar[ul]|{[1]}
&\ldots&\obj A_{n-2}\ar[ul]|{[1]}&&\obj A_{n-1}
\ar[ul]|{[1]}
} \label{eq:Pichain}
\end{equation}
where the $\obj A_j$'s are stable objects satisfying
\begin{equation}
  \xi(\obj A_0)\geq\xi(\obj A_1)\geq\ldots\geq\xi(\obj A_{n-1}).
\end{equation}
These decay chains are unique up to isomorphism \cite{Brg:stab} and
so, in particular, no decay chain (for $n>1$) may exist for a stable object.

Now, let us suppose we are working in the derived category
$\DC(Q)$. For a given point in the moduli space of complexified
K\"ahler forms, trying to determine stability of a given object using
purely the definition of $\Pi$-stability is difficult, if not
impossible. In addition, one must usually assume that the objects of
some basic collection are stable for some given set of gradings.
We will assume that the objects $L_j$ are stable for all $j$. This is
a very natural assumption given the D-brane world-volume approach to
$\theta$-stability, as outlined in \cite{DFR:stab}, but we will not
try to further justify this assumption here. For now let us also
assume that the grades satisfy the following ``partial alignment''
property:
\begin{equation}
  \psi < \xi(L_j) < \psi+1,\quad\hbox{for all $j$},
    \label{eq:align1}
\end{equation}
for some fixed real number $\psi$. We will now show that, under these
conditions, any non-trivial object which cannot be represented by a
stalk complex is necessarily unstable. Suppose we have a two-term
complex
\begin{equation}
  W^\bullet = \left(\xymatrix@1{W^{-1}\ar[r]^f&\poso{W^0}}\right).
\end{equation}
We have a morphism between complexes
\begin{equation}
\xymatrix{0\ar[r]&\ker(f)\ar[r]\ar[d]&\poso{0}\ar[r]\ar[d]&0\\
0\ar[r]&W^{-1}\ar[r]^f&\poso{W^0}\ar[r]&0,}  \label{eq:aker}
\end{equation}
which is a quasi-isomorphism if and only if $f$ is surjective. Similarly
\begin{equation}
\xymatrix{
0\ar[r]&W^{-1}\ar[r]^f\ar[d]&\poso{W^0}\ar[r]\ar[d]&0\\
0\ar[r]&0\ar[r]&\poso{\coker(f)}\ar[r]&0,}  \label{eq:acoker}
\end{equation}
is a quasi-isomorphism if and only if $f$ is injective. Therefore, in
order to avoid an isomorphism to a single term complex we should
assume that both $\ker(f)$ and $\coker(f)$ are non-trivial.

In (\ref{eq:aker}), $\ker(f)$ is a non-trivial quiver
representation. Let $j$ be the minimum number for which
$\dim(e_j\ker(f))>0$. Then there is a non-trivial quiver morphism
$L_j\to\ker(f)$. This, in turn, implies that there is non-trivial
morphism from $L_j$ to $W^{-1}$. That is, $\Hom(L_j[1],W^\bullet)$ is
nonzero. If $W^\bullet$ were stable then the unitarity constraint
(\ref{eq:unt}) would yield
\begin{equation}
  \xi(L_j)+1 \leq \xi(W^\bullet).
\end{equation}
Similarly there must exist a maximum number $k$ for which
$\dim(e_k\coker(f))>0$ yielding a map $\coker(f)\to L_k$ resulting in 
$\xi(W^\bullet)\leq\xi(L_k)$. Therefore, from (\ref{eq:align1}), we have
\begin{equation}
  \psi+1 < \xi(L_j)+1 \leq \xi(W^\bullet) \leq \xi(L_k) < \psi+1,
\end{equation}
which is absurd and so $W^\bullet$ cannot be stable.

It is a simple matter to rule out translations of the above and
complexes of length greater than two by the same argument. So we need
only consider quiver representations themselves, rather than
complexes, in order to determine the stable spectrum of objects. In
particular, if $\obj{C}$ in (\ref{eq:Pichain}) is a stalk complex then
it follows (from, for example, the Grothendieck group) that every
object in this diagram must be a stalk complex concentrated in the
same position. This turns the sequence of triangles into a sequence of
short exact sequences in the category of quiver representations. This
simplifies the analysis considerably.

Let $K_0(Q)$ be the Grothendieck group of quiver representations of
$Q$. This is simply the lattice $\Z^n$ of dimension vectors
where negative dimensions are allowed. Recall that the grading $\xi$
is determined by a central charge $Z$, where
\begin{equation}
  Z:K_0(Q)\to\C,
\end{equation}
is homomorphism. We then have 
\begin{equation}
\xi=\frac1{\pi}\arg(Z)\mod 2.
\end{equation}
It follows that, for any stable representation $V$,
\begin{equation}
  \xi(V) = \frac1{\pi}\arg\left(\sum_j e^{i\pi\xi(L_j)}\dim
  V_j\right)\mod 2.
     \label{eq:Zsum}
\end{equation}
We may fix the mod 2 ambiguity as follows. Suppose we have a stable
two-dimensional representation $V$ whose dimension vector is
$(0,0,\ldots,0,1,\allowbreak0,\ldots, 0,1,0,\ldots0)$ where the ones
appear in position $i$ and $j$ with $i<j$. If $\xi(V)<\psi-1$ then the
triangle
\begin{equation}
\xymatrix{L_j[-1]\ar[rr]&&L_i\ar[dl]\\
&V\ar[ul]|-{[1]}&
}
\end{equation}
destabilizes $L_i$, which we know is supposed to stable. Similarly, if
$\xi(V)>\psi+2$ then $L_j$ would be unstable. We therefore conclude,
using (\ref{eq:align1}), that $\psi<\xi(V)<\psi+1$. We may iterate this
process to build up any finite-dimensional representation, and so all
such representations have their gradings restricted to this unit
interval thereby resolving the ambiguity in (\ref{eq:Zsum}).

Given a short exact sequence of quiver representations
\begin{equation}
\xymatrix@1{0\ar[r]&U\ar[r]&V\ar[r]&W\ar[r]&0,}
\end{equation}
if $U$ and $W$ are stable then a necessary condition for the $\Pi$-stability
of $V$ is that $\xi(W)>\xi(U)$. From (\ref{eq:Zsum}) it is clear that
$\xi(V)$ lies between $\xi(U)$ and $\xi(W)$ and so we require
$\xi(U)< \xi(V)$. Define
\begin{equation}
  \theta(U) = -\Im\frac{Z(U)}{Z(V)}.   \label{eq:theta}
\end{equation}
This necessary condition for the stability of $V$ then becomes
$\theta(U)>0$ for all $U\subset V$.

This turns out to be a sufficient condition too as can be seen as
follows. If $\obj{C}=V$ in (\ref{eq:Pichain}) is an unstable object,
then, because the triangles have turned into short exact sequences,
$\obj{A}_0$ is subobject of $\obj{C}$. In the special case of partial
alignment we may extend the definition of grading so that it is
defined for unstable objects as well as stable objects. We simply
compute the grade of an object from (\ref{eq:Zsum}) using the
alignment to fix the mod 2 ambiguity. From this, it follows from
(\ref{eq:Pichain}) that $\xi(\obj{A}_0)\geq\xi(V)$, i.e.,
$\theta(\obj{A}_0)\leq0$.

We have arrived at precisely King's $\theta$-stability criterion of
\cite{King:th}. We have proved the following statement that
$\Pi$-stability reduces the $\theta$ stability if the grades are
sufficiently aligned:
\begin{theorem}   \label{th:Pitheta}
  If all the objects $L_j$ are stable and their grades lie within an
  interval of width one, then a necessary condition for $\Pi$-stability
  of an object in $\DC(Q)$ is that it be a stalk complex, i.e., of the
  form $V[n]$ for some quiver representation $V$. This object is then
  stable if and only if $\theta(U)>0$ for all subrepresentations
  $U\subset V$,
  where $\theta$ is defined by (\ref{eq:theta}).
\end{theorem}

The fact that $\Pi$-stability reduces to $\theta$-stability was also
proven in \cite{DFR:stab} using a D-brane world-volume field theory
argument and looking for unbroken supersymmetries. In the latter case
it is difficult to know rigorously how far away from alignment one may
go before $\theta$-stability fails (see \cite{Fiol:suN} for more on
this). We believe the above argument is more straight-forward and
clearly shows that $\theta$-stability is valid for a nonzero-sized
region in moduli space since the grades are continuous functions of
the moduli.

In the case of $\dP1$, we can now show that the condition for
$\theta$-stability coincides with the criteria laid out in theorem
\ref{th:BO}. Suppose that
\begin{equation}
  \psi < \xi(L_0) < \xi(L_1) < \xi(L_2) < \xi(p) < \xi(L_3) < \psi+1,
    \label{eq:aligndP1}
\end{equation}
where $p$ is a representation with dimension vector $(1,1,1,1)$. It is
not hard to see that $p$ will be stable if the maps in the quiver are
{\em generic}. However, if, for example, we have $b_0=b_1=0$ then
there is a nontrivial map $p\to L_1$. The kernel of this map has
dimension $(1,0,1,1)$ and is a subrepresentation of $p$ with
$\theta<0$ and thus destabilizes $p$. Similarly, the other cases of
$p$'s not corresponding to skyscraper sheaves are ruled out.

The moduli space of $\Pi$-stable objects of type $p$ is therefore the
same space as the moduli space of skyscraper sheaves and is therefore
$\dP1$ itself. Note that another choice of grades can affect this
result. For example, if
\begin{equation}
  \psi < \xi(L_0) < \xi(p) < \xi(L_1) < \xi(L_2) < \xi(L_3) < \psi+1,
    \label{eq:alignP2}
\end{equation}
then the moduli space of $p$'s is $\P^2$. This choice of grades
effectively blows the exceptional curve down.

We may generalize this result to the following
\begin{theorem} \label{th:mod}
  Assume that the objects $L_i$ are stable and that all skyscraper
  sheaves correspond to stalk complexes in $\DC(Q)$ associated to
  representations $p$ with a fixed dimension vector (given by
  (\ref{eq:skydim})). Let the nodes numbered $l,l+1,\ldots,n-1$ be the
  nodes of $Q$ on which no arrow has a head. If the grades satisfy
\begin{equation}
  \psi < \xi(L_0) < \xi(L_1) < \xi(L_2) < \ldots < \xi(L_{l-1})
      < \xi(p) < \xi(L_l) \ldots < \xi(L_{n-1}) < \psi+1,
    \label{eq:align-gen}
\end{equation}
then the moduli space of $\Pi$-stable (or, equivalently,
$\theta$-stable) objects $p$ is the del Pezzo surface $S$.
\end{theorem}
To see this one proves that the condition both for a failure of Bondal
and Orlov's condition in theorem \ref{th:BO} and the failure of
$\Pi$-stability is that there should be a node $k$, where $k<l$, to
which the representation has no nonzero paths of nonzero length. We
are assuming that the generic representation $p$ satisfies the Bondal
and Orlov condition and one shows that the projective resolution
analogous to (\ref{eq:dP1p}) can only jump to something violating the
condition, like (\ref{eq:dP1pf}) under these circumstances. The
$\Pi$-stability condition occurs since we would have a non-trivial
morphism $p\to L_k$ which induces the decay of $p$. Furthermore, if
$p$ can decay in any way, then it will be able to decay by this
channel because of the inequalities (\ref{eq:align-gen}).

Note that we have not proven that only stalk complexes can satisfy the
conditions of theorem \ref{th:BO}. However, we found this to be true for
$\dP1$ above, and it is easy enough to check for other
examples. The physics of 0-branes certainly implies that the result of
theorem \ref{th:mod} is correct and so this assumption should be true.

We may extend the above results to $\DC(\bar Q)$, i.e., the embedding
of $S$ into $X$. Once we include the dotted arrows we have
directed loops in the quiver and so we are no longer guaranteed to
have maps either to or from specific $L_j$'s. Unfortunately we needed
this property above to make the assertion that the grades of all
representations lay in the same unit interval as those of the
$L_j$'s. However, in this case one can argue that any representation of $\bar
Q$ is a deformation of a stable representation of $Q$ and so has the
same grade. This fixes the grade in the desired interval and the
remaining arguments above go through unchanged.

In particular, $\Pi$-stability for objects in $\DC(\bar Q)$ descends
to $\theta$-stability for representations of $\bar Q$. Clearly the
moduli space for points on $S$ is extended by the ability for points
to move ``off'' $S\in X$ once we pass to $\bar Q$. Thus, the moduli
space of $\theta$-stable representations of the right dimension vector
should represent the normal bundle of $S\in X$.


\section{Complete alignment of the gradings}   \label{s:align}

We now want to go to the extreme case of alignment of gradings when
\begin{equation}
  \xi(L_0)=\xi(L_1)=\xi(L_2)=\ldots=\xi(L_{n-1}).
     \label{eq:equal}
\end{equation}
Using theorem \ref{th:Pitheta} and comparing (\ref{eq:bchain}) to
(\ref{eq:Pichain}) we see that the {\em only\/} $\Pi$-stable objects
in $\DC(Q)$ are the $L_j$'s. 

What happens to the moduli space of skyscraper sheaves on $S$? It is
typical in moduli space problems to look for the set of objects in a
given class which are either themselves stable, or are a direct sum of
stable objects (``poly-stable''). This happens, for example, in the
Donaldson--Uhlenbeck--Yau theorem \cite{Don:YM,UY:YM} for the case of
Hermitian--Yang--Mills connection for which one uses
$\mu$-stability. Similarly, one may analyze a symplectic quotient
problem in the context of quivers, which was shown in
\cite{DFR:orbifold} to yield the classical moduli space of certain
fields in the D-Brane world-volume field theory. This was shown by
King \cite{King:th} to be associated to $\theta$-stability.

Again, when describing a D-Brane moduli space, one expects points in
the moduli space to be given either by stable D-Branes, or by direct
sums of stable D-branes of identical grading representing a marginally
bound state. One might worry that quantum corrections can be important
when analyzing bound states at threshold (see \cite{SS:redux}, for
example). However, we are doing a strictly classical analysis
here. Indeed, the moduli space itself is a strictly classical concept.

Skyscraper sheaves correspond to quivers whose dimension vector is
given by (\ref{eq:skydim}). Given (\ref{eq:equal}), no such quiver is
stable.\footnote{Given the analysis of exceptional sheaves in
\cite{KO:excdP}, at least three members of the exceptional collection
must have rank $\geq1$.} The only object in the moduli space of quiver
representations with this dimension vector can be the direct sum
\begin{equation}
L_0^{\rank{\cE_0}}\oplus L_1^{\rank{\cE_1}}\oplus\ldots
  \oplus L_{n-1}^{\rank{\cE_{n-1}}}.  \label{eq:Spoint}
\end{equation}
That is, the moduli space of skyscraper sheaves on $S$ is a single
point. 

Let us try to extend this result to $\DC(\bar Q)$. Adding in the
dotted arrows to the quiver, the moduli space of skyscraper sheaves is
enlarged. Associating nonzero maps to the dotted arrows excludes the
nonzero homomorphisms which gave the decomposition
(\ref{eq:bchain}). Thus, generically one would expect such D-branes
to remain stable when we impose (\ref{eq:equal}). Indeed, we would
expect points away from $S$ inside $X$ to be unaffected by our
manipulations. However, any skyscraper sheaf in $\DC(\bar Q)$
corresponding to a point on $S$ will become unstable and we replace
$S$ by a single point corresponding to (\ref{eq:Spoint}). This yields
the following
\begin{theorem}
Given the assumptions of theorem \ref{th:mod} and the genericity
assumption above, if $X_0$ represents the moduli space of (direct
sums) of $\Pi$-stable quivers in $\DC(\bar Q)$ with dimension vector
(\ref{eq:skydim}) subject to the gradings (\ref{eq:equal}) then we
have a birational map
\begin{equation}
  X\to X_0,
\end{equation}
in which the del Pezzo surface $S$ is contracted to a point.
\end{theorem}
In other words, totally aligned gradings make $S$ collapse to a point.
Conversely, a generic skyscraper sheaf in $\DC(Q)$ will be stable if
$\xi(L_i)\leq\xi(L_j)$ for all $i<j$ unless we have the strict alignment
(\ref{eq:equal}). In this sense $S$ will {\em only\/} shrink down to a point
if we have complete alignment.

Such a contraction has been observed directly by computing periods for
the case of the $S=\P^2$ in \cite{DG:fracM,DFR:orbifold} and
$S=\P^1\times \P^1$ in \cite{AM:delP}. Similarly, it was proven in
\cite{me:TASI-D} that the periods align at any orbifold point. Note
that in these latter cases, $X$ is defined as the target space of the
string $\sigma$-model whereas our general view in this paper is that
$X$ is the moduli space of 0-branes. We will not attempt here to delve
into the profound question of whether these two points of view are
equivalent!

It is precisely when the grades are perfectly aligned that the open
strings between the D-branes $L_j$ form the field content of a quiver
gauge theory (see \cite{AM:delP}, for example). Thus, the quiver gauge
theory description (unperturbed by the addition of Fayet--Iliopoulos
terms) is relevant when, and only when, the del Pezzo surface
collapses to a point.

So far we have always assumed $\xi(L_i)\leq\xi(L_j)$ for $i<j$. The extreme
converse case $\xi(L_0)>\xi(L_1)>\ldots>\xi(L_{n-1})$ represents a
situation where the moduli space of skyscraper sheaves becomes
completely empty. In such cases the embedding of $\DC(S)$ into $\DC(X)$
becomes geometrically meaningless. Generally one can then describe the
geometry by a different embedding $\DC(S')\subset\DC(X)$ of another
del Pezzo surface $S'$ (which might be equivalent to $S$) into
$X$. This may represent some kind of flop transition, or a
re-identification of $\DC(S)$ under some quantum symmetry of $X$.


\section*{Acknowledgments}

I wish to thank M.~Douglas, I.~Melnikov, D.~Morrison, R.~Plesser
and M.~Stern for useful conversations. The author is
supported in part by NSF grants DMS-0074072 and DMS-0301476.


\end{document}